# Two-step flux synthesis of ultrapure transition metal dichalcogenides

Song Liu[1*#], Yang Liu[1*#], Luke Nemetz Holtzman[2], Baichang Li[1], Madisen Holbrook[1], Jordan Pack[3], Takashi Taniguchi[4], Kenji Watanabe[4], Cory R. Dean[3], Abhay Pasupathy[3], Katayun Barmak[2], Daniel A. Rhodes[1&], James Hone[1*]

[1]Department of Mechanical Engineering, Columbia University, New York, New York 10027, USA

[2]Department of Applied Physics and Applied Mathematics, Columbia University, New York, New York 10027, USA

[3]Department of Physics, Columbia University, New York, New York 10027, USA

[4]National Institute for Materials Science, Tsukuba, Ibaraki 305-0044, Japan

[#]These authors contributed equally to this work.

[&]Present address: Department of Materials Science and Engineering, University of Wisconsin, Madison, WI, USA

[*]Email: songliu0128@gmail.com; liuyang.nano@gmail.com; jh2228@columbia.edu

**Two-dimensional transition metal dichalcogenides (TMDs) have attracted tremendous interest due to the unusual electronic and optoelectronic properties of isolated monolayers and the ability to assemble diverse monolayers into complex heterostructures. To understand the intrinsic properties of TMDs and fully realize their potential in applications and fundamental studies, high-purity materials are required. Here, we describe synthesis of TMD crystals using a two-step flux growth method that eliminates a major potential source of contamination. Detailed characterization of TMDs grown by this two-step method reveals charged and isovalent defects with densities an order of magnitude lower than in TMDs grown by a single-step flux technique. For WSe$_2$, we show that increasing the Se:W ratio during growth reduces point defect density, with crystals grown at 100:1 ratio achieving charged and isovalent defect densities below $10^{10}$ cm$^{-2}$ and $10^{11}$ cm$^{-2}$, respectively. Initial temperature-dependent electrical transport measurements of monolayer WSe$_2$ yield room-temperature hole mobility above 840 cm$^2$/Vs and low-temperature disorder-limited mobility above 44,000 cm$^2$/Vs. Electrical transport measurements of graphene-WSe$_2$ heterostructures fabricated from the two-step flux grown WSe$_2$ also show superior performance: higher**

**graphene mobility, lower charged impurity density, and well-resolved integer quantum Hall states. Finally, we demonstrate that the two-step flux technique can be used to synthesize other TMDs with similar defect densities, including semiconducting 2H-MoSe$_2$ and 2H-MoTe$_2$, and semi-metallic T$_d$-WTe$_2$ and 1T'-MoTe$_2$.**

The ability to observe and harness new physical phenomena depends on the synthesis of high-purity materials and devices, with III-V heterostructures (i.e., GaAs) being the most prominent example[1]. Two-dimensional (2D) van der Waals heterostructures (vdWHs) have likewise seen similar improvements in performance and emergence of new phenomena with decreasing disorder[2-4]. For exfoliated graphene, which has total defect densities below ~$10^{10}$ cm$^{-2}$,[5] encapsulation within hBN dramatically reduces environmental disorder and improves electronic behavior.[6,7] Transition metal dichalcogenides (TMDs)[8], on the other hand, have been limited by a high density of point defects reaching ~$10^{13}$ cm$^{-2}$.[2,9,10] Thus, synthesis of high-purity TMDs is the key obstacle to realizing the promise of these materials. Thin film synthesis is very important for applications but still shows very large defect density[11-14]. Single crystals grown by chemical vapor transport (CVT) are commonly used as a source to exfoliate flakes for basic studies but can also suffer from very high defect density (Supplementary Figure 1)[15,16]. We have previously shown that a self-flux synthesis technique yields crystals with lower point defect density and improved performance, but the total defect density is still at least two orders of magnitude higher than that of graphene or hBN[15,17,18].

Here we report the use of a two-step flux synthesis technique to eliminate a key source of contamination within flux-grown TMDs, using WSe$_2$ as a model TMD. We use scanning tunneling microscopy (STM) to identify two classes of defects (charged and isovalent) in flux-grown WSe$_2$, and develop a protocol for measuring the density of each type. We find that the two-step synthesis technique leads to a dramatic reduction in point defect density and seek to optimize the process by varying the Se:W ratio during synthesis. We find that the density of charged and isovalent defects both decrease with increasing ratio, reaching values of order $3\times10^9$ and $8\times10^{10}$ cm$^{-2}$, respectively, at a ratio of 100:1. Initial electrical transport measurements of monolayer WSe$_2$ show a room-temperature hole mobility exceeding 840 cm$^2$/Vs and low-T mobility exceeding 44000 cm$^2$/Vs. As a further test of the electronic purity of the WSe$_2$, we assembled a graphene-WSe$_2$ heterostructure in which the graphene acts as a proximate sensor for charged impurities in monolayer WSe$_2$. At high magnetic fields, the flux-derived device shows well-resolved integer quantum Hall states, comparable to devices encapsulated in hBN.

**Conventional Single-Step Self-flux Synthesis and Defects Counting**

We have previously reported the use of a conventional single-step self-flux synthesis process to grow crystals of WSe$_2$ and MoSe$_2$[15,19]. In this method, transition metal (M) and chalcogen (X) materials are sealed with a quartz wool filter inside a quartz ampoule under high vacuum (see Supplementary Figure 2). The sample is heated to 1080 °C and slowly cooled (~2 °C/h) so that crystals precipitate out of the M-X solution. At an intermediate temperature (~400 °C), the ampoule is removed from the oven and the crystals are separated from the melt by centrifugation through the quartz wool (additional residue is removed by two-zone annealing). This strategy is widely used and previous literature has not cited any adverse effects of the quartz wool on TMD crystal quality[20,21].

In order to assess the quality of TMD crystals accurately, we employ scanning tunneling microscopy (STM) to directly image point defects. To avoid surface contamination, each crystal is cleaved *in situ* in ultrahigh vacuum (UHV). Figure 1 shows images of the surface of a WSe$_2$ crystal grown by the single-step self-flux technique. Importantly, we find two classes of defects that occur with different frequency and require imaging at different magnification to ensure accurate quantification. A large-area (500×500 nm$^2$) scan (Fig. 1a) shows a large number of prominent bright spots whose contrast reverses when the bias is inverted (Supplementary Figure 3), indicating that these defects are charged. A small-area (50×50 nm$^2$) scan (Fig. 1b) of an area free of the charged defects reveals a second class of defect. The contrast for these defects does not reverse when the bias is inverted, indicating that they are isovalent. Atomic-resolution images of these two defect types, highlighting the larger spatial extent of the charged defect, are shown in Figs. 1c and 1d. In the following sections, we simply refer to these two classes of defects as charged and isovalent defects.

In order to determine the density of charged defects, we count the number appearing within five different large-area scans taken over different areas of the crystal. Error bars for all reported defect density values represent the statistical variation across multiple scans. In the crystal shown in Fig. 1a, the charged defect density is $(1.3 \pm 0.14) \times 10^{11}$ cm$^{-2}$. This value is consistent with our previous study, and well below the defect density found in CVT-grown crystals[15]. For isovalent defects, we average 30 small-area scans. In the crystal shown in Fig. 1b, the isovalent defect density is $(1.44 \pm 0.22) \times 10^{12}$ cm$^{-2}$. We note that this defect counting method may over-count defects on a single layer, since those lying below the first monolayer may also appear in the STM images; previous STM defect measurements on bulk TMDs have shown that point defects can be sensed in the top 3 layers[15]. However, any over-counting (by up to a factor of 3) will not affect the trends observed below with changes in the synthesis recipe.

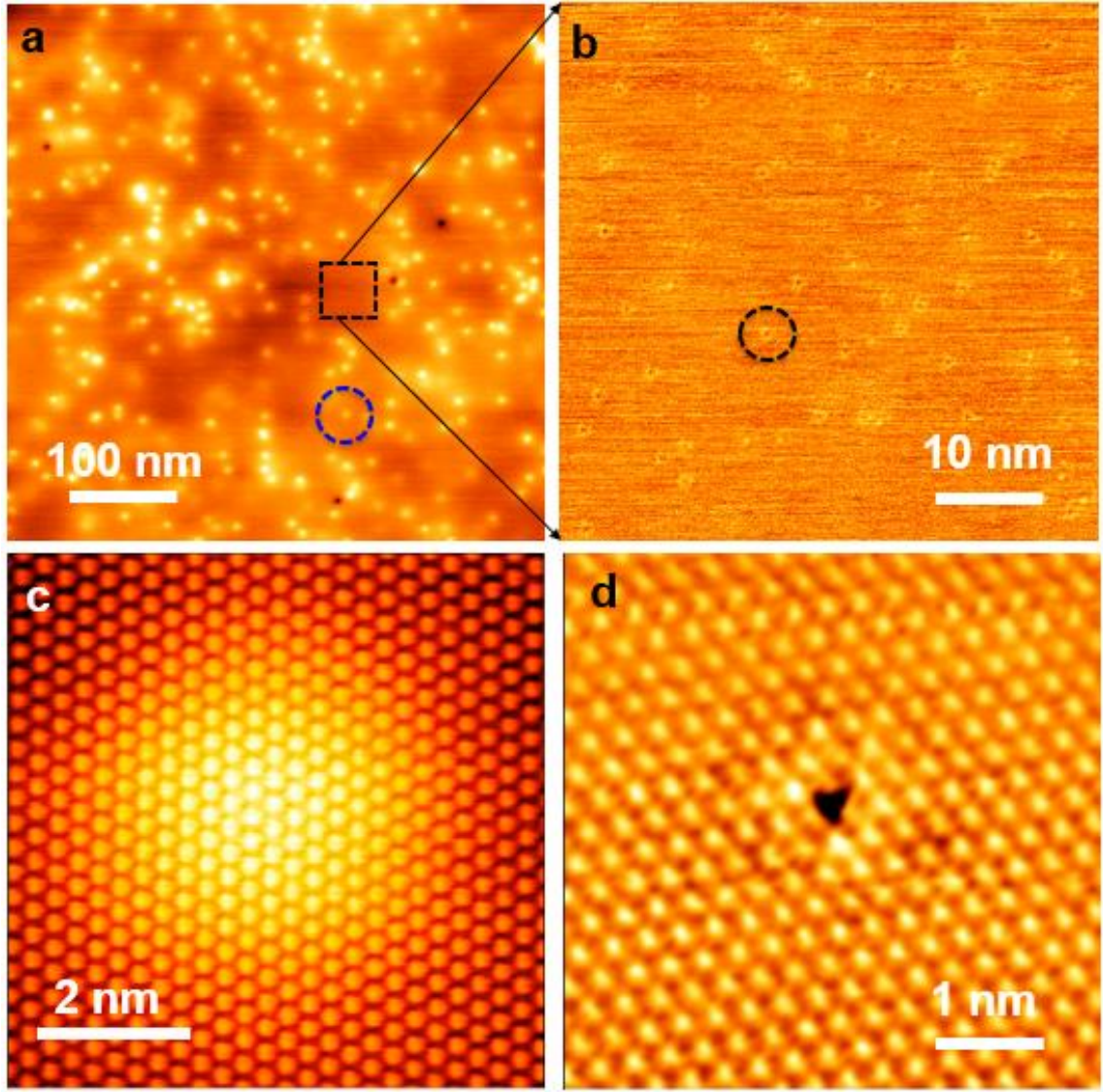

**Figure 1 | STM imaging of single-step flux-grown WSe$_2$.** STM topographic images of WSe$_2$ crystal synthesized with the conventional single-step flux method, with 5:1 Se:W ratio. **(a)** 500×500 nm$^2$ image showing charged defects as bright and dark points. **(b)** 50×50 nm$^2$ image of the area in the black dashed square in (a). **(c)** High-resolution image of the charged defect circled in (a). **(d)** High-resolution image of

the isovalent defect circled in (b). The parameters for room-temperature STM imaging are $V = 1.4$ V and $I = 200$ pA.

Although more work is needed to determine the precise nature of the charged and isovalent defects, we hypothesized that the quartz wool might act as a source of impurities: even though the wool is not in physical contact with the melt, evaporation of impurities from the high-surface-area wool under the high-temperature (~1000 °C), high pressure (~20 atm) synthesis conditions is certainly plausible[22,23]. A common alternative to the use of quartz wool is to use an alumina crucible with integrated frit (Canfield crucible)[24]; however, the high pressure in this process requires thick-walled quartz tubing that makes the use of alumina crucibles difficult. Therefore, we developed a two-step flux method (Fig. 2a) that separates the crystal synthesis process from the post-process removal of excess chalcogen. For the synthesis process, transition metal (M) and chalcogen (X) materials (with excess chalcogen, *i.e.* X:M molar ratio above 2) are sealed in a quartz ampoule under high vacuum (~$10^{-6}$ Torr). After heating and dwelling at high temperature (1000-1150 °C) for 2 weeks, each ampoule is slowly cooled down (1 °C/h) to 600 °C and then quickly cooled to 250 °C. At this point, we slowly cool the $MX_2$ crystals to allow enough time for the solid selenium to delaminate from the quartz ampoule. From here, the $MX_2$ crystals are embedded within solid chalcogen. To separate the $MX_2$ crystals from the chalcogen, we use a post-process step, where again quartz wool is utilized as a filter inside a second vacuum-sealed ampoule. This post-process step only requires brief heating to the melting point of the chalcogen, greatly reducing the contamination from the quartz wool. Finally, the filtered crystals are loaded into a third vacuum-sealed ampoule and annealed in a temperature gradient to remove the remaining chalcogen residue (see Methods for details).

STM imaging of a $WSe_2$ crystal grown using the two-step process with a 5:1 Se:W ratio shows a dramatic reduction in the density of charged defects (Fig. 2b) and isovalent defects (Fig. 2c) compared to a crystal grown at the same 5:1 ratio using the single-step process (Figure 1). Electrical transport measurements can provide an important measure of the quality of the two-step flux-grown crystals. However, we have found that achieving low-temperature Ohmic contact to high-purity semiconducting monolayer TMDs is particularly challenging due to the relatively large bandgaps and lack of doping due to defects. Recently, we have developed a technique to achieve robust p-type contacts through selective hole-doping of the contact regions via charge transfer from RuCl$_3$[25]. Initial measurements of the temperature-dependent mobility of a WSe$_2$ monolayer encapsulated within hBN are shown in Figure 2d (see details in Methods). At a hole density of $3\times10^{12}$ cm$^{-2}$, the room-temperature mobility is 840 cm$^2$/Vs. This is comparable with the theoretical phonon-limited mobility[26]. At low temperature, where mobility is limited by scattering from disorder, the mobility exceeds 44,000 cm$^2$/Vs. This is the largest reported value for a monolayer TMD to date and attests to the extremely high quality of the crystal[27]. Detailed studies of the transport and magneto-transport behavior are ongoing and will be reported elsewhere.

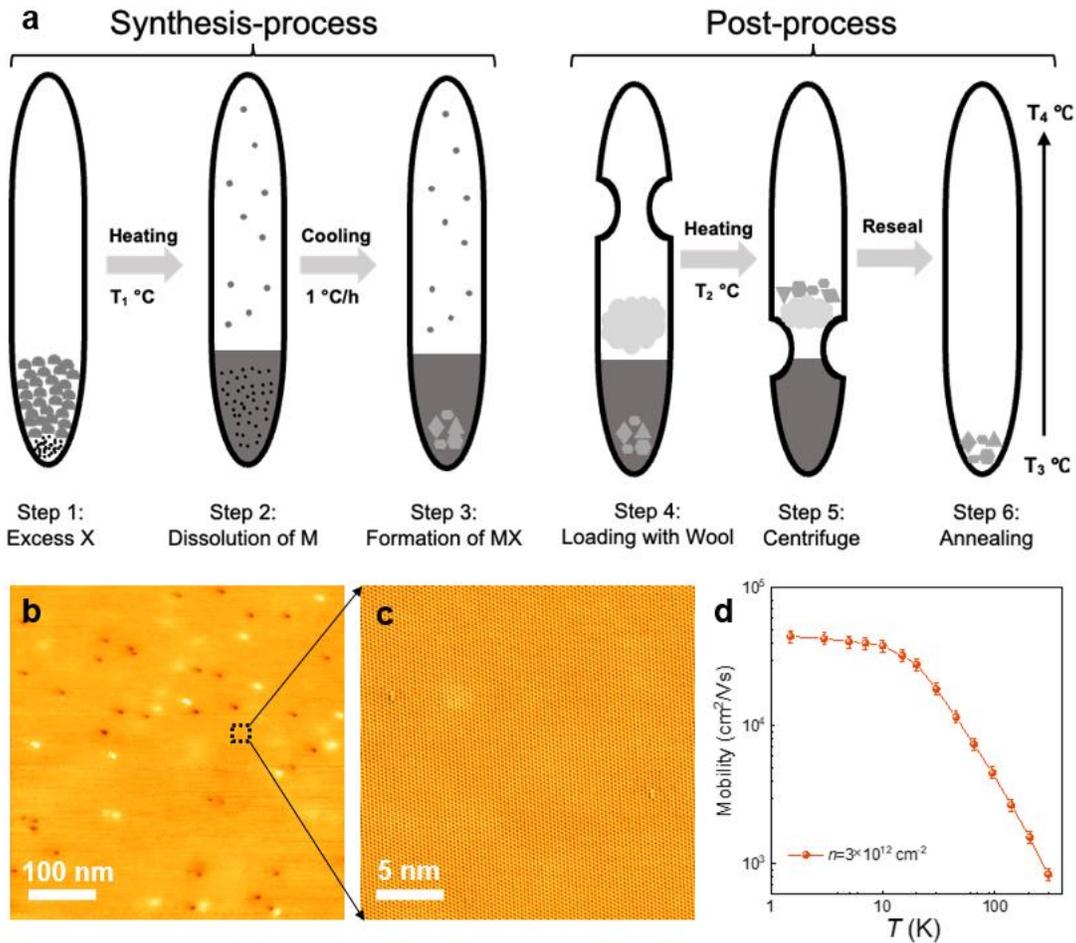

**Figure 2 | Two-step flux synthesis. a**, Schematics of two-step flux synthesis technique. (**b-c**) STM topographic images of WSe$_2$ crystals synthesized with two-step flux method at different magnification level. The X:M ratio is 5:1. The parameters of STM imaging are *V*=1.4 V and *I*=200 pA. (d) Temperature-dependent hole mobility of monolayer WSe$_2$. Measured mobility of a Hall bar device consisting of hBN-encapsulated monolayer WSe$_2$ at hole density of $3\times10^{12}$ cm$^{-2}$.

We next studied how the Se:W ratio affects the density of point defects. We used the two-step flux process to synthesize WSe$_2$ from melts with various Se:W ratios of 3:1, 20:1, and 100:1. Representative large-area and small-area STM images of the resulting crystals are shown in Figure 3. These clearly demonstrate that the density of both charged and isovalent defects decreases with increasing Se concentration. Figure 3g shows the charged and isovalent defect density obtained at each Se:W ratio. Notably, the 100:1 recipe achieves the lowest defect densities with a charged defect density of $(3.8 \pm 1.8) \times 10^9$ cm$^{-2}$ and an isovalent defect density of $(8.1 \pm 4.7) \times 10^{10}$ cm$^{-2}$. Figures 3h and 3i demonstrate the statistical results of defect density with a narrow distribution, highlighting the robustness of this two-step flux method. To quantify the reproducibility of the synthesis and defect characterization technique, we measured defect density from three crystals grown in different ampoules with the same Se:W ratio and experimental conditions. As shown in Figs. 3h and 3i, we find that the charged defect density generally stays below the $10^{10}$ cm$^{-2}$ level, and the isovalent defect concentration is less than $3 \times 10^{11}$ cm$^{-2}$.

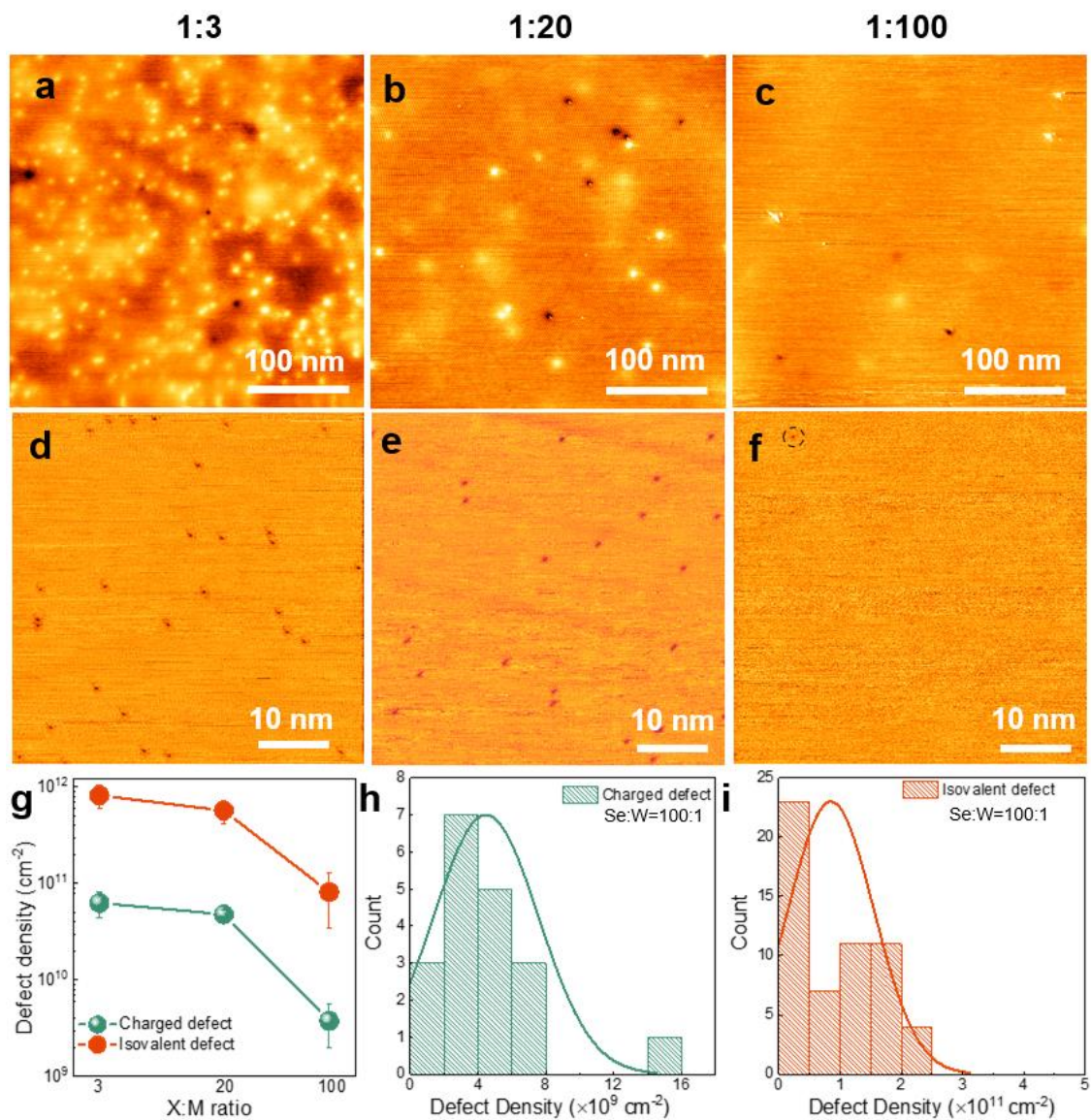

**Figure 3 | Effect of Se:W ratio on defect density. (a-f),** STM topographic images of two-step flux-grown WSe$_2$ with Se to W ratios at 3:1 (**a, d**), 20:1 (**b, e**) and 100:1 (**c, f**) at different magnification. **g,** Defects density versus X: M ratio; Error bars are obtained from sample-to-sample variation across different crystals grown in the same batch. (**h-i**), Statistical defects density of charged (**h**) and isovalent (**i**) defect density.

**WSe$_2$ as a low-disorder substrate for graphene**

Semiconducting TMDs have been identified as potential substrates for graphene that can exhibit lower disorder than amorphous oxides while offering a more promising route to scalable production than hBN[28]. In addition, TMDs can be used to induce spin-orbit coupling in bilayer graphene, as a route to achieving topological properties[29]. However, defects in TMDs can induce charge disorder, diminish graphene mobility, and introduce hysteresis. Recent work[30] has shown that devices with graphene encapsulated in superacid-treated $WS_2$ show minimal hysteresis, reduced charge disorder ~$10^{11}$ cm$^{-2}$, and room-temperature mobility at or near the acoustic phonon limit. Here, we use the same methodology to establish that the $WSe_2$ crystals described above can serve as an ultralow-disorder substrate. Figure 4 shows the behavior of a sample using monolayer $WSe_2$ derived from the highest-purity (Se:W=100:1 ratio) two-step flux-grown crystal, with the configuration shown in Figure 4a (see details in Methods)[31,32]. For comparison, a second sample was fabricated using CVT-grown $WSe_2$, with corresponding data shown in Supplementary Figure 4. Figure 4b shows the measured resistivity as a function of gate voltage at room temperature and low temperature. The charge neutrality point is within 2 V of zero gate voltage, indicating minimal doping (~$10^{11}$ cm$^{-2}$). The room-temperature mobility reaches the acoustic phonon limit (matching that of an equivalent hBN-graphene-hBN device) for electron or hole density above $3.5 \times 10^{12}$ cm$^{-2}$, while dropping to slightly lower values at lower doping densities (Supplementary Figure 4). Figure 4c shows the temperature-dependent mobility of the graphene at a hole density of $2\times10^{12}$ cm$^{-2}$. Its low-temperature mobility of ~$1.6\times10^5$ cm$^2$/Vs corresponds to mean free path of ~1.85 µm. This value matches the device size, indicating that mobility is limited by scattering from edges rather than by impurities. By contrast, the CVT device shows roughly 8 times lower mobility at low T and 2.8 times lower mobility at room temperature. Figure 4d shows the measured low-temperature conductivity as a function of electron density. On this plot, the red lines show the intersection point between the constant conductivity at low density and the linear behavior at high density, which provides a good measure of the charge inhomogeneity ($n^*$) in the graphene[33]. The derived value of $n^* = 1.4 \times 10^{10}$ cm$^{-2}$ is much smaller than that achieved in the CVT device and an order of magnitude lower than that reported for superacid-passivated $WS_2$[30]. This value is well below the concentration of defects identified as isovalent by STM (Fig. 3 above), confirming that these defects have negligible contribution to the charge inhomogeneity in the graphene. The measured $n^*$ is somewhat larger than the charged defect density identified by STM. However, the underlying $SiO_2$ substrate can also give rise to charge disorder of order $10^{10}$ cm$^{-2}$ hBN-encapsulated graphene devices[2]. Ongoing studies using graphite gates, which can reduce environmental charge disorder into the $10^9$ cm$^{-2}$ range[2], will more sensitively probe charge disorder arising specifically from the $WSe_2$. In this data, we also observe a plateau in the doping range of $3.5-7\times10^{11}$ cm$^{-2}$. We hypothesize that this plateau arises from the filling of defect (trap) states near the conduction band of the $WSe_2$. Its width corresponds to a trap density of $2.5\times10^{11}$ cm$^{-2}$, consistent with the isovalent defect density observed by STM (Fig. 3 above). A much more prominent feature is observed in the same vicinity for the CVT sample, (see Supplementary Figure 4) further supporting this interpretation.

We next examine whether the high-purity WSe$_2$ can be used as a substrate for quantum transport measurements. Figure 4e plots the graphene conductivity as a function of both carrier density and applied field. Well-developed half-integer quantum Hall features at $\nu = \pm 2, 6, 10$ *etc*. appear below 2 T, and broken-symmetry states ($\nu = \pm 1, 3, 4, 5$ *etc*.) appear at high fields, particularly in the hole-doped region (Supplementary Figure 4g). We note that a subtle distortion of the fan diagram is seen at $n \sim 3\times10^{11}$ cm$^{-2}$, matching the plateau seen in Fig. 4d that we assign to trap states. The CVT sample shows a much more prominent distortion of the fan diagram at the same density, consistent with higher trap state density. The presence of trap states near the conduction band of WSe$_2$ may also account for the lower quality of the fan diagram on the electron side compared to the hole side.

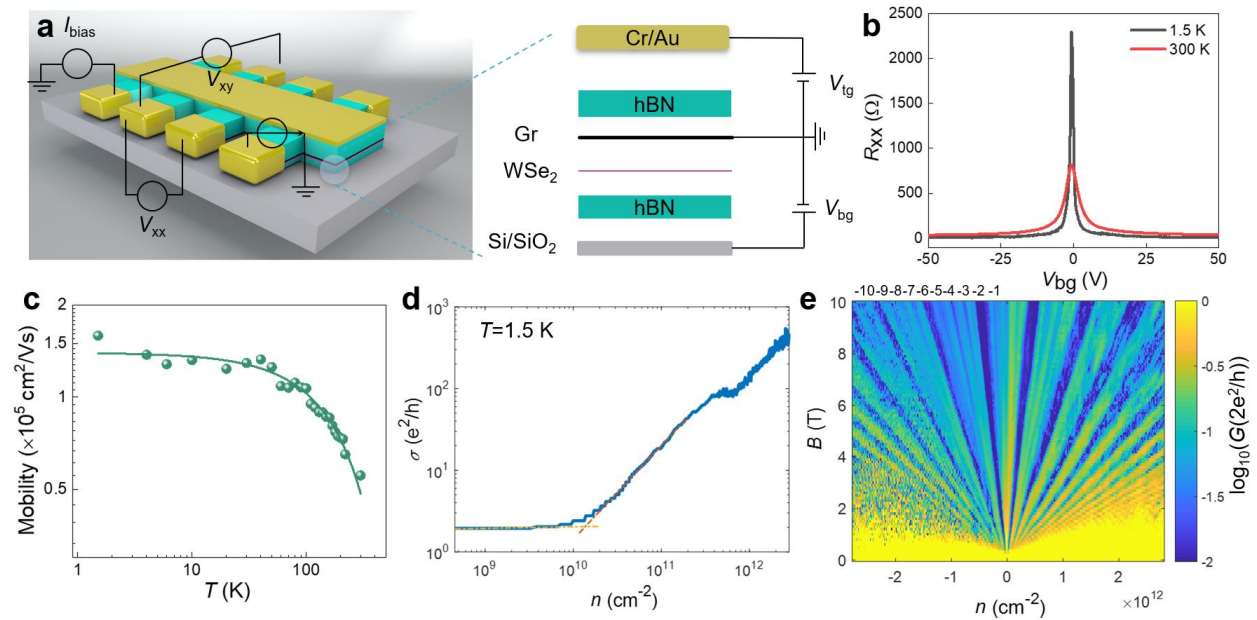

**Figure 4 | Electrical transport behavior of graphene on WSe$_2$.** (**a**) Schematic device structure and cross-section. The two hBN layers are ~40 nm thick, while the graphene and WSe$_2$ are monolayers. The WSe$_2$ monolayer is derived from a crystal grown by the two-step flux synthesis method at 100:1 Se:W ratio. The substrate is Si with 285 nm SiO$_2$. (**b**) Measured $R_{xx}$ vs. back-gate voltage ($V_{bg}$) at room temperature and low temperature (1.5 K). (**c**) Graphene mobility as a function of temperature, at hole density of $2 \times 10^{12}$ cm$^{-2}$. (**d**) Conductivity as a function of carrier density. Dashed lines show extrapolation of the constant low-

density conductivity and the linear dependence at higher density. **(e)** Landau fan diagram of the flux-TMD device. Integer filling fractions are indicated at top.

**General Synthesis of ultraclean TMDs**

The two-step flux synthesis technique can readily be extended to other selenide and telluride-based-TMDs (sulfides are more challenging due to high sulfur vapor pressure). Figure 5 shows STM images of two semiconducting TMDs (2H-MoSe$_2$ and 2H-MoTe$_2$) and two semi-metallic TMDs: (T$_d$-WTe$_2$ and 1T'-MoTe$_2$) grown at 100:1 X:M ratio (see Methods for details). Charged defects are circled in the large-area images (Figs. 5a-5d), and all samples show charged defect density below $2\times10^{10}$ cm$^{-2}$. 2H-MoSe$_2$ and 2H-MoTe$_2$, and T$_d$-WTe$_2$ are all obtained by slow cooling from the growth temperature. On the other hand, 1T'-MoTe$_2$ is obtained by rapid quenching from 900 °C to room temperature. This rapid quenching does not seem to increase the charged defect density. Figures 5e and 5f show high-resolution STM images and fast Fourier transforms of the 2H and 1T' phases of MoTe$_2$, highlighting the change in crystal symmetry.

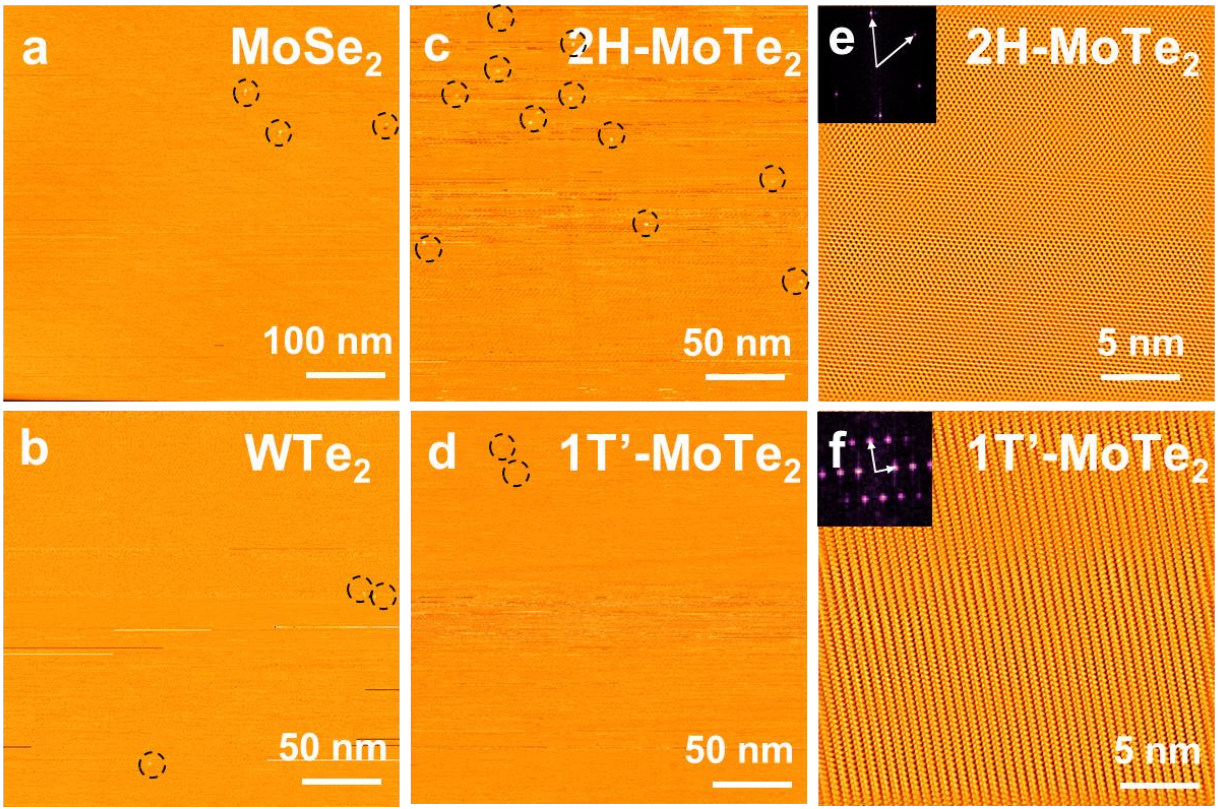

**Figure 5 | General synthesis of different TMDs.** STM topographic images of 2H-MoSe$_2$ (**a**), T$_d$-WTe$_2$ (**b**), 2H-MoTe$_2$ (**c**) and 1T'-MoTe$_2$ in (**d**) grown using the two-step flux synthesis method. Charged defects

are circled. (**e,f**) high-resolution STM images of 2H and 1T' phases of MoTe$_2$, showing the different crystal symmetry. Insets are corresponding FFT plots.

**Conclusion**

To summarize, we have demonstrated a general two-step flux synthesis method for growing high-purity Se- and Te-based TMDs with low defect density. This method reduces the density of charged and isovalent defects by roughly one order of magnitude when compared to a single-step process. The defect density decreases when the chalcogen-metal ratio within the flux increases, reaching ~$10^9$ cm$^{-2}$ for charged defects and ~$10^{11}$ cm$^{-2}$ for isovalent defects. Initial electrical transport measurements of high-purity WSe$_2$ show record-high room-temperature and low-temperature mobilities of above 840 cm$^2$/Vs and 44,000 cm$^2$/Vs, respectively. Graphene-WSe$_2$ heterostructures show outstanding performance with two-step flux-derived WSe$_2$, including high mobility, low charged impurity density, and well-resolved integer quantum Hall states. Finally, we extend our two-step flux synthesis technique, enabling the synthesis of other semiconducting and metallic transition metal selenides and tellurides with high purity. This breakthrough in high-purity materials synthesis sets the stage for the next round of advances towards realizing novel quantum states in TMDs and towards applications in electronics and optoelectronics.

**Methods**

**Growth of Se-based crystals.**

M powders (99.9975% Mo or 99.999% W) and excess selenium shot (99.999+%) were loaded at specific ratio (X:M) into a quartz ampoule and sealed at high vacuum (~$10^{-6}$ Torr). These ampoules were then gradually heated to 1000 ℃ over 48 h, held at 1000 ℃ for two weeks, and subsequently cooled at a rate of 1 °C/h to 600 °C before quickly cooling to 250 °C. At 250 °C, the ampoules were cooled over 1 day to room temperature to allow the selenium to delaminate from the quartz as it solidified. At this stage, the crystals and excess Se were transferred to a new ampoule with an additional quartz wool for filtering excess Se and sealed under vacuum. The ampoule was heated to 285 ℃ and then centrifuged to separate the crystals from the excess Se. Finally, the MSe$_2$ crystals were transferred to a third vacuum-sealed ampoule and subjected to a temperature-gradient ($T_{\text{hot}} = 260$ ℃, $\Delta T \sim 100$℃) with the crystals placed at the hot end for 16 h, then cooled to room temperature.

**Growth of Te-based crystals.**

M powders (99.9975% Mo or 99.999% W) and excess tellurium broken ingot (99.9999+%) were mixed with a specific ratio (X:M) into a quartz ampoule and sealed at high vacuum (~$10^{-6}$ Torr). These ampoules were gradually heated to 1100 °C over 48 h and held at 1100 °C for one week. For 1T'-MoTe$_2$, the samples were cooled at a rate of 1 °C/h to 880 °C and then abruptly quenched in water to maintain the 1T' phase. For 2H-MoTe$_2$ and T$_d$-WTe$_2$, the samples were cooled at a rate of 1 °C/h to 550 °C before quenching to room temperature in air. For the second (filtering) step, the temperature was 550 °C (2H-MoTe$_2$ and T$_d$-WTe$_2$) or 900 °C (1T'-MoTe$_2$) to liquefy the solid Te. Finally, the MTe$_2$ crystals were transferred to a third vacuum-sealed ampoule and subjected to a temperature-gradient ($T_{\text{hot}} = 420\ °C, \Delta T \sim 200°C$) with the crystals placed at the hot end for 16 h and then cooled to room temperature.

**STM characterization.**

STM measurements were performed using a Scienta Omicron STM system at room temperature under an ultra-high vacuum (base pressure lower than $1\times10^{-10}$ torr). Bulk TMD crystals were mounted onto a metallic sample holder with silver epoxy and then cleaved *in situ* in the UHV STM chamber to obtain a clean surface. Before all measurements, the tungsten tip was cleaned and calibrated against an Au(111) surface. Next, the defect density was calculated for each STM image by dividing the number of defects by the scanning size. To avoid uncertainty from local inhomogeneity while counting, each value for the defect density was obtained by averaging several STM images across the sample surface.

**Electrical device fabrication and measurement.**

Flux-grown WSe$_2$ and hBN bulk crystals were mechanically exfoliated using the Scotch tape technique. Monolayer WSe$_2$ and few-layer hBN were exfoliated onto 285 nm SiO$_2$/Si substrates and then identified via optical microscopy. The thickness of the flakes was finally confirmed by a combination of Raman microscopy (Renishaw Raman system) and AFM (Bruker Atomic Force Microscope) measurements.

The hBN/graphene/monolayer WSe$_2$/hBN was assembled layer-by-layer using a dry-stacking technique with polypropylene carbonate (PPC)[32]. The monolayer WSe$_2$ device was also assembled using the dry-stacking method, but instead with polycarbonates (bisphenol A) (PC), where RuCl$_3$ was used as the doping layer for the contact regions. The final device geometry was defined by electron-beam lithography (EBL) and reactive ion etching (RIE, Oxford Plasmalab 100 ICP-RIE instrument). For electrical edge contacts to the hBN/graphene/monolayer WSe$_2$/hBN device, Cr/Au=2/100 nm was deposited via e-beam evaporation. The monolayer WSe$_2$ device utilized graphite contacts with RuCl$_3$ as the contacts doping layer, following the procedure in Ref.[34]. Transport measurements were performed using a standard low-frequency lock-in amplifier with an excitation frequency of 17.777 Hz and an excitation current bias of 100 nA (using a 10 MΩ resistor) in a dry-cryostat with temperatures ranging from 1.5 K to 300 K.

The device parameters of hBN/graphene/monolayer WSe$_2$/hBN: the width of the Hall bar is 2 $\mu$m, and the length between two $R_{xx}$ voltage bars are 1.25 $\mu$m. The device parameters of hBN/monolayer WSe$_2$/hBN device: the hall bar is 1 $\mu$m wide and the spacing between voltage leads is 1.5 $\mu$m.

**Acknowledgments:**

This work was primarily supported by the NSF MRSEC program at Columbia through the Center for Precision-Assembled Quantum Materials (DMR-2011738). Transport studies were supported by the Department of Energy (DE-SC0016703). Synthesis of boron nitride (K.W. and T.T.) was supported by the Elemental Strategy Initiative conducted by the MEXT, Japan (grant no. JPMXP0112101001), and JSPS KAKENHI (grant nos. JP19H05790 and JP20H00354).

**Author Contributions:** S.L., Y.L., K. B., D.A.R. and J. H. conceived and designed the project. S. L. and Y. L. contributed equally to this work. S. L. and D. A. R. developed and optimized the flux growth strategy, S. L., Y. L., and L. N. H. grew the TMDs crystals. Y. L. and B. C. L. performed the heterostructure stacking. Y. L. carried out nanofabrication, electrical measurement, and data analysis. S. L. and M.A.H. performed the STM measurements. J. P. and C. D. made and measured the monolayer WSe$_2$ device. K. W. and T. T. provided hBN crystals. J. H., Y. L. and S. L. co-wrote the manuscript with input from all authors. All authors contributed to writing and discussing of the manuscript.

**Competing Financial Interests:** The authors declare no competing financial interests.

**Data Availability**: The data that support the findings of this study are available from the corresponding authors upon reasonable request.